\documentclass[12pt]{article}

\title{Statistical Mechanics of\\
       Multiply Wound D-Branes}
\author{Gavin Polhemus\thanks
       {Electronic address: {\tt g-polhemus@uchicago.edu}}}
\date{\emph{Department of Physics and Enrico Fermi Institute\\
            University of Chicago\\
            5640 Ellis Avenue, Chicago, Illinois 60637}}
\newcommand{\EFInumber}{EFI-96-48}
\newcommand{\archivenumber}{hep-th/9612130}

\newcommand{\eg}{{\it e.g. }}
\newcommand{\mean}[1]{\left\langle #1 \right\rangle}
\newcommand{\abs}[1]{\left| #1 \right|}
\newcommand{\commute}[2]{\left[ #1,#2 \right]}

\newcommand{\infinity}{\infty}
\newcommand{\textfrac}[2]{{\textstyle\frac{#1}{#2}}}
\newcommand{\partialwith}[1]{\frac{\partial}{\partial#1}}
\newcommand{\sth}{^{\mbox{\scriptsize th}}}
\newcommand{\Tr}{\mathop{\rm Tr}\nolimits}
\newcommand{\diag}[1]{\mathop{\rm diag}\nolimits\{#1\}}

\newcommand{\eqref}[1]{(\ref{#1})}
\newcommand{\hepref}[3]{#2, ``#3,'' {\tt hep-th/#1}.}
\newcommand{\hepjref}[4]{#2, ``#3,'' #4, {\tt hep-th/#1}.}
\newcommand{\hepitem}[3]{\bibitem{#1}\hepref{#1}{#2}{#3}}
\newcommand{\hepjitem}[4]{\bibitem{#1}\hepjref{#1}{#2}{#3}{#4}}
\newcommand{\NPB}[3]{{\sl Nucl. Phys. }{\bf B#1} (#2) #3}
\newcommand{\PLt}[3]{{\sl Phys. Lett. }{\bf #1B} (#2) #3}

\newcommand{\prob}{\mathcal{P}}
\newcommand{\partition}{\mathbf{Z}}

\newcommand{\greekdef}[3]
 {\def#1{\relax\ifmmode#2\else#3\fi}}

\def\b{\beta}

\newcommand\z{\zeta}

\greekdef\o{\theta}{\char"1C}
\greekdef\i{\iota}{\char"10}

\greekdef\l{\lambda}{\char32l}
\newcommand\m{\mu}
\newcommand\n{\nu}
\newcommand\x{\xi}
\newcommand\p{\pi}
\def\r{\rho}
\newcommand\s{\sigma}

\newcommand\D{\Delta}
\greekdef\O{\Theta}{\char"1F}
\greekdef\L{\Lambda}
 {\leavevmode\setbox0\hbox{L}\hbox to\wd0{\hss\char32L}}

\greekdef\P{\Pi}{\mathhexbox27B}
\greekdef\S{\Sigma}{\mathhexbox278}

\begin{document}

\begin{titlepage}

\maketitle
\thispagestyle{empty}

\begin{table}[t!]
\rightline{\EFInumber}
\rightline{\archivenumber}
\end{table}

\begin{abstract}
The D-brane counting of black hole entropy is commonly understood in 
terms of excitations carrying fractional charges living on long, 
multiply-wound branes (\eg open strings with fractional Kaluza-Klein 
momentum).  This paper addresses why the branes become multiply wound.  
Since multiply wound branes are T-dual to branes evenly spaced around 
the compact dimension, this tendency for branes to become multiply 
wound can be seen as an effective repulsion between branes in the 
T-dual picture.  We also discuss how the fractional charges on 
multiply wound branes conspire to always form configurations with 
integer charge.
\end{abstract}

\end{titlepage}

\section{Introduction}
\setcounter{footnote}{0}

The past year has seen considerable progress in understanding black 
hole entropy through counting states of D-brane configurations.  This 
approach has been particularly fruitful in the study of 
five-dimensional black holes \cite{9601029}.  One of the most 
carefully studied models is that of black holes in the toroidal 
compactification of type IIB supergravity down to five dimensions 
\cite{9602043}.  When discussing the applications of the ideas below, 
this model will be used as an example.  These five dimensional black 
holes are characterized by three charges, two asymptotic fields and 
mass.  Since it is not known how to calculate the entropy of these 
configurations directly, the black hole is modeled by a configurations 
of 5-branes, 1-branes, their anti-branes and strings carrying the same 
charges.  In certain limits (all near BPS) the entropy of the D-brane 
configurations can be calculated by directly by counting microstates.  
In these cases the entropy of the D-brane configuration has been found 
to agree with the Beckenstien-Hawking entropy for a black hole with 
the same charges, fields and mass.  Further calculations have shown 
agreement in the absorption cross sections, and low energy emission 
spectra \cite{9605234,9606185,9609026,9612051}.

In order for the D-brane to give the right entropy in the limit of 
large charges and mass the configurations must have excitations that 
carry fractional charges.  These fractional charges also lead to the 
large density of string and small mass gap that are expected for a 
black hole.  This has been commonly understood as an indication that 
the branes which wrap compact dimensions, rather than forming a large 
collection of singly wound branes, join together to form a single, 
long, multiply-wound brane \cite{9601152,9604042,9605016}.

The notion of D-brane winding is not as obvious as fundamental string 
winding.  When $Q$ D-branes coincide, the gauge theory on the brane is 
enlarged to a $U(Q)$ symmetry.  If the brane wraps a compact dimension 
($x_{9}$ in this example), then this gauge symmetry can be broken by 
adding a Wilson line $A^{9} = \diag{\o_{1},\o_{2},\ldots, \o_{Q}}/2\p 
R_9$ \cite{9602052}.  This leads to a holonomy given by the matrix
\begin{equation}
 U_9=\pmatrix
  {e^{i\o_{1}} & 0           & \cdots & 0          \cr
   0           & e^{i\o_{2}} &        & 0          \cr
   \vdots      &             & \ddots & \vdots     \cr
   0           & 0           & \cdots & e^{i\o_{Q}}\cr
  }.
\end{equation}
In the T-dual picture (which we will refer to as the ``unwound 
picture'' to distinguish it from the ``wound picture'' above) this is 
a set of parallel branes at positions $R_9(\o_{1},\o_{2},\ldots, 
\o_{Q})$ around the compact dimension.

Of course we could also add Wilson lines that have off diagonal 
elements.  For example, one case that represents a single, 
multiply-wound brane is
\begin{equation}
 U_9=\pmatrix
  {0      & 1 & 0      & \cdots & 0      \cr
   0      & 0 & 1      &        & 0      \cr
   \vdots &   & \ddots & \ddots & \vdots \cr
   0      & 0 &        & 0      & 1      \cr
   1      & 0 & \cdots & 0      & 0      \cr
  }.
 \label{woundmat}
\end{equation}
However, one can always put such a matrix in diagonal form by a change 
of basis.  So the Wilson lines with off diagonal elements break the 
gauge symmetry \emph{exactly} like the diagonal Wilson line with the 
same eigenvalues.  For example the eigenvalues of (\ref{woundmat}) are 
the $Q\sth$ roots of one.  So a multiply wound brane is T-dual to 
parallel branes that are evenly spaced around the compact dimension.  
As described in \cite{9602052,9610250}, this leads to fractionally 
charged excitations.  However, the total charge of a configuration is 
always integer.  This restriction is discussed in Section 
\ref{integer} below.

The issue of how branes become multiply wound so that they can carry 
fractional charges can be address by considering how the ``unwound'' 
D-branes spread out around the compact dimension.  If the branes were 
non-interacting they could be expected to spread out to fill the 
compact dimension much the way particles of a non-interacting gas 
spread out to fill a container.  However, we show in this paper that 
the branes do not behave this way and in fact spread out to fill the 
compact dimension with with remarkable uniformity.

\section{How Branes Repel}

The most obvious reason for the branes to spread out evenly around the 
compact dimension would be to minimize some sort of repulsive 
potential.  However, there is absolutely no way for a potential to be 
generated---Wilson lines do not break the supersymmetries so, in the 
absence of excitations, all Wilson lines must have zero energy.  
Likewise, in the unwound picture there are no forces pushing the 
branes apart because any distribution of parallel branes preserves 
some of the supersymmetries and therefore has zero energy.

Since all Wilson lines are zero energy, we assume that all Wilson 
lines are equally probable when doing D-brane thermodynamics.  It may 
be that the process of creating the D-brane configurations leads to 
certain Wilson lines more frequently than others, but that would be a 
rather exceptional situation.  Certainly, one can carefully prepare a 
specific configuration, but in a realistic scenario we anticipate that 
interactions during preparation lead to every zero energy 
configuration with equal probability.  Note that, since all of the 
states are zero energy, the temperature of the system does not matter.

While all Willson lines are equally probable, not all positions of the 
branes are equally probable!  There is only one element of $U(Q)$ with 
all its eigenvalues equal to 1, but there are many whose eigenvalues 
are, for example, the $Q\sth$ roots of one.  The probability that the 
branes will have positions $(\o_{1},\o_{2},\ldots, \o_{Q})$ around the 
compact dimension is proportional to the volume of $U(Q)$ which 
eigenvalues $(e^{i\o_{1}},e^{i\o_{2}},\ldots, e^{i\o_{Q}})$.

\subsection{An Example with Two Branes (Not 2-Branes)}

As an example, consider the case of two coincident branes.  In this 
case the gauge group is $U(2) = U(1) \times SU(2)$.  The $U(1)$ 
subgroup is the center of mass position in the unwound picture, so the 
branes relative position is given by the eigenvalues of the $SU(2)$.  
The elements of $SU(2)$ can be parameterized by $\x^j$ using the three 
Pauli matrices, $\s_j$, as generators:
\begin{equation}
	U = e^{i\x^j \s_j}.
\end{equation}
The eigenvalues are then $e^{\pm i|\x|}$, resulting in a separation 
between the branes $\o_1 - \o_2 = 2|\x|$.  Notice that $SU(2)$ is 
topologically $S^{3}$ and the surfaces of constant $|\x|$ are like the 
spheres of constant latitude running from $|\x| = 0$ at the north pole 
to $|\x| = \p$ at the south pole on $S^3$.  The radius of each sphere 
is $r = \sin{|\x|}$ so the probability that the two branes are at 
positions $\o_1$ and $\o_2$ is
\begin{equation}
	\prob (\o_1,\o_2) = \sin^{2}\left( \frac{\o_1 - \o_2}{2} \right).
\end{equation}
The probability density actually vanishes for the branes being 
coincident.  The peak of the probability distribution is at maximum 
separation in the unwound picture, which corresponds to a doubly wound 
brane in the wound picture.

\subsection{Many Branes}
\label{Many Branes}

The probability of a randomly chosen element of $SU(Q)$ having 
eigenvalues $(\o_1,\o_2,\ldots\o_Q)$ has been exactly solved for all 
$Q$ \cite{Random}, 
\begin{equation}
 \prob (\o_1,\o_2,\ldots,\o_Q) = 
  C \prod_{k < l}^Q \left| e^{i\o_k} - e^{i\o_l} \right|^2 .
 \label{Dbrane}
\end{equation}	
The maximum is clearly at equal spacing.  
Fixing the origin at $\mean{\o_Q} = 0$ and putting the remaining $\o_j$ 
in increasing order gives $\mean{\o_j} = 2\p j / Q$.

To gain an intuitive understanding of the branes' behavior it is 
useful to consider another physical system that has the same 
probability distribution: a collection of $Q$ particles, each with 
charge $q$, living in two dimensions on a circle of radius $R$ 
\cite{Random,9604037}.  The two dimensional world can be thought of as 
the complex plane.  The two-dimensional Coulomb potential for a pair 
of charges at $z_k$ and $z_l$ is $W_{kl} = -q^2 \ln|z_k - z_l|$.  
Restricting the charges to the unit circle, $z_j = R e^{\o_j}$.  The 
energy of the system is then
\begin{equation}
 W = -q^2 \sum_{k<l}^{Q} \ln \left| e^{i\o_k} - e^{i\o_l} \right|.
\end{equation}

In a thermal ensemble of such systems at inverse temperature $\b$ 
the probability distribution will be
\begin{eqnarray}
 \prob (\o_1,\o_2,\ldots,\o_Q) 
  &=& C e^{-\b W} \\
  &=& C \prod_{k < l}^Q
   \left| e^{i\o_k} - e^{i\o_l} \right|^{\b q^2}. 
\end{eqnarray}	
For $\b = 2/q^2$ this is exactly the probability distribution in 
(\ref{Dbrane}).  Note that the inverse temperature $\b$ has nothing to 
do with the temperature of the branes or the black hole.  This is the 
temperature of the model system of electric charges that has the same 
probability distribution as the D-brane configuration at any 
temperature.

Clearly the charges are going to have a very strong tendency to spread 
out evenly around the circle, forming a sort of crystal.  In the case 
of only two particles it is easy to see that the fluctuations in the 
spacing are roughly the same as the average spacing, $\D(\o_1-\o_2) 
\approx \mean{\o_2-\o_1} = \p$.  It turns out that as the number of 
particles gets large and the average spacing gets smaller, the 
fluctuations in nearest neighbor spacing get proportionally smaller so 
that $\D(\o_j-\o_{j+1}) \approx \mean{\o_j-\o_{j+1}} = 2\p/Q$ 
\cite{Random}.

Although the spacings between adjacent particles varies significantly, 
long wavelength fluctuations in the average spacing are minuscule.  
These long wavelength fluctuations can be studied in the continuum 
limit.  In this case the discrete index, $j$, is replaced by a 
continuous parameter, $\x = 2\p j/Q$.  The positions can be expanded 
about their means in Fourier modes,
\begin{equation}
 \o_\x = \mean{\o_\x} 
  + \sum^{Q-1}_{n \neq 0} a_n e^{in\x},
\end{equation}
where $a_{-n} = a_n^*$ and $\mean{\o_\x} = \x$.
The energy of the system is the ground state energy plus the energies 
of these modes,
\begin{equation}
 W = W_0 + 
  \sum^{Q-1}_{\scriptstyle n = -(Q-1) \atop \scriptstyle n \neq 0}
   W_n.
\end{equation}
Calculating these energies of these fluctuations is a simple exercise 
in two dimensional electrostatics:
\begin{equation}
 W_n = 2\p^2 \r_0^2 n |a_n|^2,
\end{equation}
where $\r_0$ is the average charge density, $\r_0 = Qq/2\p$.

As the number of charges gets large all of the long wavelength 
fluctuations actually freeze out.  This is best seen by fixing the 
average charge density of the system, $\r_0$.  Since we are interested 
in the temperature at which the charged particles are distributed like 
the ``unwound'' D-branes, $\b = 2/q^2 = Q^2 / 2\p^2 \r_0^2$, the 
temperature decreases like $1/Q^2$.

To calculate the average energy in these modes at $\b = 2/q^2$ we use 
the partition function,
\begin{eqnarray}
 \partition_n 
  &=& \int_0^\infinity da_n \; e^{-\b W_n} 				\\
  &=& \frac{1}{2 \r_0 \sqrt{2\p \b n}} .
\end{eqnarray}
The mean energy in these modes is then
\begin{equation}
 \mean{W_n} = -\partialwith{\b}\ln\partition_n	= \frac{1}{2\b}.
\end{equation}
From this we can easily find the 
average (root mean squared) amplitude of these macroscopic oscillations,
\begin{eqnarray}
 \mean{|a_n|^2} 
  &=& \frac{1}{(2\p\r_0)^2 n\b }				\\
  &=& \frac{1}{2 Q^2 n}.
\end{eqnarray}

This can be used to find the fluctuations in the spacing between 
remote particles,
\begin{eqnarray}
 \mean{(\o_\x-\o_\z)^2} 
  &=& \mean{\biggl[ \x-\z + 
   \sum_{n \neq 0} 
    a_n \left( e^{in\x} - e^{in\z} \biggr) 
     \right]^2}									\\
  &=& (\x - \z)^2 
   + \sum_{n > 0}
    \mean{\abs{a_n}^2} 
    \left( e^{in\x} - e^{in\z} \right)
    \left( e^{-in\x} - e^{-in\z} \right)	\\
  &=& \mean{\o_\x - \o_\z}^2 
   + \frac{1}{Q^2} 
   \sum_{n > 0} \frac{1}{n} 
    \sin^2 \frac{n (\x -\z)}{2} .
\end{eqnarray}
The infinite sum is divergent and needs to be cut off at $n=Q/2$.  The 
sum can be approximated by an integral; the sine cuts off the integral 
in the infrared at $n = \p / (\x-\z)$.
\begin{equation}
 \sum_{n > 0} \frac{1}{n} 
  \sin^2 \frac{n (\x -\z)}{2}
 \approx \frac{1}{2}
  \int_{\frac{\p}{(\x-\z)}}^{\frac{Q}{2}} \frac{dn}{n}
 = \textfrac{1}{2} \ln (k-l) ,
\end{equation}
where we have used $\x = 2\p k/Q$ and $\z = 2\p l/ Q$.  So the size of 
the fluctuations in spacing for $k-l \gg 1$ is 
\begin{eqnarray}
 \D(\o_k - \o_l) 
  &\approx& \frac{1}{Q} \sqrt{\textfrac{1}{2} \ln (k-l)}.
\end{eqnarray}
As we would expect, the fluctuations in spacing get bigger as we 
consider particles that are farther apart.  What is surprising is how 
slowly the fluctuations increase.

Since the probability distribution for the particles is the same as 
that for the unwound branes, all of the above statements about 
deviations from equal spacing applies to the brane case as well.  
However, the mechanism is different.  The particles spread out to 
minimize their energy while the branes spread out to find larger 
volumes of phase space.  Because of this difference, the charged 
particles cannot be used to model all aspects of the branes' 
behavior---for example, the momentum distribution of the branes in 
different.

All interesting D-brane models of black holes have more than one 
compact dimension, each with its own (possibly trivial) Wilson line.  
These Wilson lines all transform together under the $U(Q)$ symmetry, 
allowing a trace term in the Lagrangian that gives non-zero energy to 
configurations with Wilson lines that cannot be simultaneously 
diagonalized \cite{9510135}:
\begin{equation}
 V = \frac{T^2}{2} \sum_{\m,\n = D}^{9} \Tr \commute{X^\m}{X^\n}^2,
 \label{V}
\end{equation}
where $T$ is the fundamental string tension and $D$ is the dimension 
of the non-compact space-time.
Both the additional Wilson lines and the 
potential that come from the additional compact dimensions may be 
important in understanding D-brane models.  However, D-brane 
calculations of black hole entropy have only required that the branes 
become multiply wound around a single $S^1$ \cite{9604042}.
Based on this we expect that the remaining four compact dimensions 
do not affect this winding.

\section{Integer Winding and Momentum of D-Brane Configurations}
\label{integer}

Non-extremal configurations have non-zero energy, allowing excitations 
away from zero potential in \eqref{V}.  These excitations 
correspond to loops of unexcited fundamental string connecting the 
various branes \cite{9510135}.  
Since winding is easier to visualize than momentum, we 
start by looking at the fractional winding states that connect the 
branes in the unwound picture.  T-Duality allows us to apply this 
understanding to the fractional momentum states in the wound D-brane 
picture.

In the unwound picture the excitations are strings connecting the 
D-branes.  Since there are a lot of branes around, the strings don't 
have to go all the way around the compact direction---they are allowed 
to have fractional winding.  However, the entire configuration must 
have integer winding.  This is a result of charge conservation on the 
branes.  The charge on the brane that comes from the end of a string 
must be canceled by an opposite charge from another string.  In terms 
of the orientation of the strings: each brane must have as many 
strings beginning on it as it has ending on it.  This guarantees that 
the net winding of any configuration will be zero.

T-duality turns the fractional winding that appeared in the unwound 
picture into fractional momentum in the wound picture.  The 
excitations of wound branes are free to move independently.  However, 
their ends are still charged, so the excitations trail field 
lines in the D-brane as they move.  Moving a single excitation around 
the compact direction wraps the field lines, changing the energy 
of the system.  Since moving a single excitation around the compact 
dimension does not restore the system to its original configuration, 
single excitations don't see the periodicity of the compact dimension 
and there is no reason for their momentum to be quantized in units of 
$1/R$.

To avoid trailing any field lines, excitations must band together into 
parties that are neutral in the D-brane theory.  These parties can 
then circumnavigate the compact dimension and discover its 
periodicity.  It is the momentum of these parties which is quantized 
in units of $1/R$.  It is easy to verify that the sum of the 
fractional momenta that make up a neutral party is always integer.

\section{Discussion}

In the models of five-dimensional black holes the excitations are 
strings connecting D-1-branes to D-5-branes.  We expect that the 
``repulsion'' described above encourages both the 1-branes and the 
5-branes to link up 
into long, multiply-wound branes, allowing fractionally charged 
excitations.  (The total charge will still be integer for the reasons 
given above)

It may be possible to understand models of near extremal 
black holes entirely in the language D-branes (without 
any fundamental strings attached) by replacing the fundamental strings 
with excitations away from $V=0$ in \eqref{V}.
However, since the fundamental strings in black hole models connect branes 
of different dimensionality, it is less clear how to proceed.

It would also be interesting to see if these ideas could be extended 
to the branes of M-theory.

\section{Acknowledgments}

I would like to thank Jeff Harvey for many enlightening discussions on 
many matters and Juan Maldacena for a helpful conversation about 
charge conservation on branes.  This work was supported by an Office 
of Naval Research Graduate Fellowship.


\end{document}